\documentclass[pre,superscriptaddress,twocolumn,floatfix,longbibliography]{revtex4-1}

\usepackage{graphics}
\usepackage{graphicx}
\usepackage{amsmath}
\usepackage{amssymb,xcolor}
\usepackage{soul}

\newcommand{\textin}[1]{\mbox{\scriptsize{#1}}}

\definecolor{grisclair}{rgb}{0.6,0.6,0.6}

\newcommand{\beq}{\begin{equation}}
\newcommand{\ee}{\end{equation}}

\begin{document}

\title{Impulsive ejection of Worthington jets by flow focusing. Influence of a soluble surfactant}
\author{T. González-Rubio}
\author{S. Rodríguez-Aparicio}
\author{E. J. Vega}
\author{J. M. Montanero}
\address{Depto.\ de Ingenier\'{\i}a Mec\'anica, Energ\'etica y de los Materiales and\\ 
Instituto de Computaci\'on Cient\'{\i}fica Avanzada (ICCAEx),\\
Universidad de Extremadura, E-06006 Badajoz, Spain}

\begin{abstract}
We experimentally investigate impulsive Worthington jet generation by flow focusing, considering the liquid meniscus location that maximizes the flow-focusing effect. This effect increases the jet velocity by up to two orders of magnitude over the bottom liquid velocity after impact. At sufficiently high liquid velocities, jet formation is governed primarily by inertia, making jet velocity largely independent of liquid viscosity, surface tension, and surfactant effects. Each liquid exhibits an optimum impact velocity that maximizes momentum transfer. Surfynol 465 lowers the impact-velocity threshold for water-jet emission and increases jet velocity by rapidly restoring equilibrium surface tension. In contrast, Triton X-100 has an almost negligible effect because of its slower adsorption kinetics. Importantly, both surfactants reduce jet diameter and, consequently, droplet size. This reduction is particularly significant with Surfynol 465 at high impact velocities, suggesting potential applications, for instance, in high-resolution printing technologies. The jet loaded with Surfynol 465 was even faster than that of the glycerine-water mixture with the same Ohnesorge and Weber numbers. This suggests that adding Surfynol 465 further accelerates jet emission, exceeding the speed attained by a clean interface with the same reduced equilibrium surface tension.
\end{abstract}

\maketitle

\section{Introduction}

% Worthington jet. Different ways of producing a Worthington jet
Impulsive and highly focused liquid jets have attracted considerable interest owing to their potential applications in several areas such as needle-free drug delivery and printing technologies, particularly for the controlled deposition of highly viscous fluids \citep{baxter2006,delrot2016,onuki2018,gordillo2020,kamamoto2021}. These transient jets undergo intense axial stretching and eventually eject a tiny droplet from their tip \citep{ZKFL00,ABLV07}. Their ability to concentrate liquid into small volumes moving at high velocities makes these jets particularly attractive for depositing highly viscous fluids and delivering precise drug doses through the skin \citep{onuki2015,onuki2018,tagawa2013,schoppink2022}. Such jets can be generated through several mechanisms, including the impulsive acceleration of a liquid column \citep{FrancoGomez2021,Krishnan2022} followed by flow focusing \citep{onuki2015,onuki2018,gordillo2020}, the impact of a drop on a solid surface \citep{BJB06}, light-scattering-induced liquid motion \citep{SWCZD07}, the collapse of a cavity following bubble bursting at a free surface \citep{G17a,LED18}, and the rapid laser-induced vaporization of liquid, which generates pressure waves that drive the surrounding fluid and produce high-speed microjets \citep{tagawa2012,schoppink2022}.

% Impulsive jets
The generation of highly focused liquid jets following different types of impulsive acceleration has been investigated in several studies. Early work by \citet{birkhoff1948} and \citet{milgram1969} examined liquid-jet formation following cavity collapse and vertical impact, respectively. Subsequent studies by \citet{ABLV07} and \citet{bergmann2008} investigated interface dynamics under impulsive disturbances and the mechanisms leading to the formation of narrow liquid jets. \citet{tagawa2012} showed that flow focusing after cavity collapse can produce highly focused, supersonic microjets. The spatiotemporal evolution of focused liquid jets following impulsive acceleration has been described by \citet{gordillo2020}.

% Impulsive jets by flow focusing. 
Worthington jets \citep{Worthington1882} can be generated by flow-focusing the momentum produced by an impulsive impact on a concave gas–liquid interface in a thin tube \cite{onuki2015}. This mechanism produces jets with velocities up to two orders of magnitude higher than the initial impact velocity in the container. In this way, extremely viscous liquids can be injected without requiring solvents, heating, or UV treatment \cite{onuki2018}. 
 
% Surfactants
Surfactants adsorb at the interface, modifying their interfacial mechanical properties \citep{A16,MS20,NED21,PPS22,VM24,M24}, even at trace concentrations \citep{RodriguezAparicio2026}. At higher concentrations, they can drastically alter multiphase flows at millimeter and submillimeter scales, affecting droplet coalescence \citep{OJ19,CBKSCJM21} and fragmentation \citep{KWTB18,PRHEM20}, bubble stability \citep{RVCMLH24,HLFCM25}, bubble-bursting dynamics \citep{CKBSCJM21,VM24}, and tip streaming \citep{A16,M24}.

% Fast-kinetics surfactants
Surfactant adsorption kinetics play a key role in determining transient interfacial properties during rapid flows \citep{VSQCC24,MS20,QCSSCC25,CKSCJMCC23,FVMGF26}. When a new interface is created, for instance during the growth of a Worthington jet, its initial surface tension is close to that of the pure solvent and decreases progressively as surfactant molecules diffuse and adsorb. The characteristic adsorption timescale is governed by molecular diffusion, the interfacial adsorption energy barrier, or both \citep{MS20,PRHEM20,CKSCJMCC23}. This timescale often exceeds that of interface creation. Standard surfactants, despite substantially reducing equilibrium surface tension, are generally too slow to replenish the interface. Uneven surfactant distribution produces Marangoni stress, leading to partial interface immobilization. Only ultrafast surfactants can significantly and homogeneously reduce surface tension during rapid events. 

% This paper
In this work, we experimentally investigate impulsive Worthington jet generation by flow focusing. We consider the liquid meniscus location that maximizes the flow-focusing effect. We will show that each liquid exhibits an optimum impact velocity that maximizes momentum transfer. At high impact velocities, inertia dominates jet formation, making it largely independent of liquid properties and surfactants. However, surfactants can affect both the impact velocity threshold for jet emission and the momentum transferred to the jet at low and moderate impact velocities. These effects depend on adsorption kinetics: Surfynol 465 lowers the jetting threshold and enhances jet velocity at low impact velocities, whereas Triton X-100 has little influence. Both surfactants reduce the size of the first emitted droplet, particularly Surfynol 465 at high velocities.

\section{Formulation of the problem}
\label{for}

\subsection{Impulsive generation of jets by flow focusing}
\label{for1}

\begin{figure}
\begin{center}
\resizebox{0.8\linewidth}{!}{\includegraphics{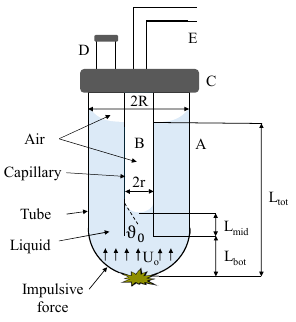}}
\caption{Device used to generate the jet: tube (A), capillary (B), plug (C), port used to fill the tube with the working liquid (D), and port connected to a syringe pump (E). A plastic rod coupled to an electrodynamic shaker hits the bottom of the tube to trigger the liquid motion.}
\label{setup1}
\end{center}
\end{figure}

% Microfluidic device
Figure \ref{setup1} shows the device used to generate the jet. A thin capillary with inner radius $r$ is positioned concentrically inside a cylindrical test tube with inner radius $R$, at a distance $L_{\mathrm{bot}}$ from the tube bottom. The tube is partially filled with a liquid with density $\rho$, viscosity $\mu$, and surface tension $\sigma$. The pressure inside the capillary is controlled to fix the heights $L_{\mathrm{mid}}$ and $L_{\mathrm{tot}}$ of the inner and outer meniscus, respectively. The inner meniscus contact angle at equilibrium is $\theta_0$, while $\theta_a$ and $\theta_r$ are the advancing and receding contact angles, respectively.

\begin{figure}
\begin{center}
\resizebox{0.8\linewidth}{!}{\includegraphics{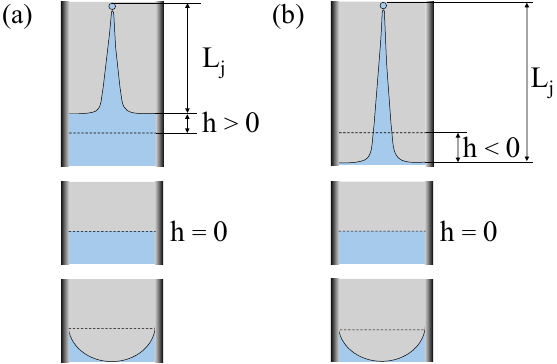}}
\caption{Phases of the Worthington jet emission. From bottom to top, (a) the meniscus bottom rises with the triple contact line pinned to the solid surface until the interface becomes flat, and (b) the Worthington jet forms with the contact line moving either upward or downward. The final position of the contact line can correspond to $h>0$ or $h<0$.}
\label{ilus}
\end{center}
\end{figure}

% The relevant quantities
An impact on the tube bottom generates an impulsive excitation that moves the meniscus bottom and produces the interface reversal (the meniscus becomes flat), giving rise to a Worthington jet (Fig.\ \ref{ilus}) \citep{onuki2018}. The liquid velocity at the tube bottom, $U_0$, averaged over the time between impact and meniscus reversal (see the Supplemental Material), measures the energy injected into the liquid before jet formation. We will focus on the jet length $L_j$ just before the ejection of the first droplet, and the average velocity $V_j=L_j/t_j$, where $t_j$ is the time between jet formation (meniscus reversal) and the ejection of the first droplet (Fig.\ \ref{ilus}). 

% The case Lmid=0
According to \citet{onuki2018}, the jet velocity for a clean liquid can be estimated as
\begin{equation}
\frac{V_j}{U_0}=\beta\frac{L_{\textin{tot}}-L_{\textin{bot}}}{L_{\textin{mid}}},
\end{equation}
where $\beta$ is a function of the $\theta_0$ and the Reynolds and Weber numbers (see Sec.\ref{dn}). This suggests that, for a given $U_0$, the velocity $V_j$ is maximum at $L_{\mathrm{mid}}=0$. We have verified this in our experiments (see the Supplemental Material). Hereafter, we focus on this singular case, for which the analysis of \citet{onuki2018} does not apply.

\subsection{Phases of the jet ejection}
\label{im}

% Phenomenon
The jet ejection for $L_{\mathrm{mid}}=0$ proceeds as follows. An impact on the test tube bottom generates an impulsive excitation on a timescale of order $10^{-4}$ s or less (see the Supplemental Material). The liquid accelerates below the meniscus, which induces flow focusing. Flow focusing moves the meniscus bottom upward and produces the interface reversal (the meniscus becomes flat), giving rise to a Worthington jet on a timescale of order 0.1 s. Specifically, the jet emission results from three consecutive phases (Fig.\ \ref{ilus}): (i) the deformation of the meniscus until the meniscus becomes flat ($\theta\simeq 90^{\circ}$) with the triple contact line pinned to the surface, (ii) the formation of the Worthington jet with the contact line either remaining still or moving upward (contact line advancing, $\theta>\theta_a$) depending on the liquid and the impact velocity, and (iii) the jet growth with the contact line moving either upward (contact line advancing, $\theta>\theta_a$) or downward (contact line receding, $\theta<\theta_r$) depending on the liquid and the impact velocity. 

When the droplet is ejected, the contact line position can correspond to $h>0$ or $h<0$ (Fig.\ \ref{ilus}). While the flow is inviscid during the impact-induced impulsive motion, viscous effects may become significant during flow focusing because the boundary layer can spread over distances on the order of $r$. In addition, surface tension is expected to constitute a significant sink of momentum for sufficiently small Weber numbers.

\begin{figure}
\begin{center}
\resizebox{0.8\linewidth}{!}{\includegraphics{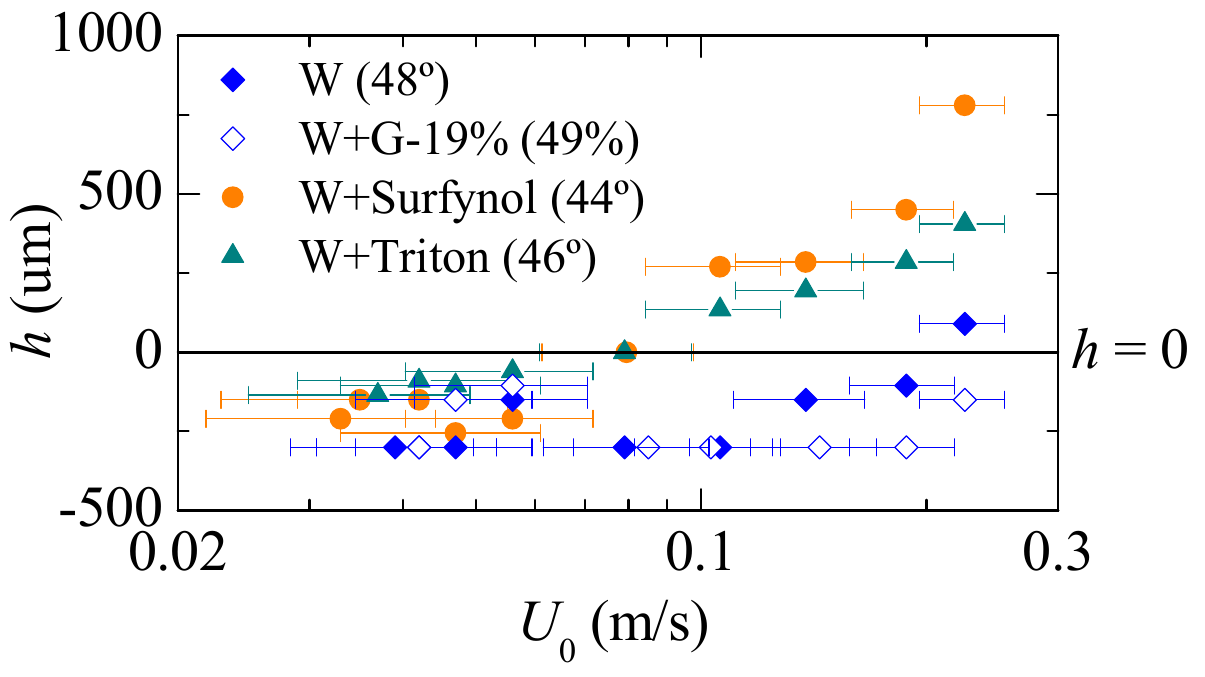}}
\caption{Triple contact line position $h$ at the droplet ejection.}
\label{ilu2}
\end{center}
\end{figure}

% Final value of h
Interestingly, without surfactant, the final contact-line position $h$ is negative in practically all cases, independent of the impact velocity (Fig.\ \ref{ilu2}). Conversely, with surfactant, $h>0$ for sufficiently large impact velocity. In other words, adding surfactant lets the liquid spread over the inner capillary surface as the faster jets grow. This resembles droplet spreading after impact on a smooth surface, where adding surfactant enhances this mechanism \citep{pasandidehfard1996,VSQCC24}. Surfynol 465 allows the meniscus to reach higher $h$ values, likely because of its faster kinetics. As shown in Sec.\ \ref{res}, $h$ is much smaller than the jet length $L_j$, suggesting that meniscus dynamics play a secondary role in jet growth. However, meniscus dynamics may affect the energy focusing on the jet, resulting in a significant variation of the momentum transferred to the jet. We will return to this aspect of the problem in Sec.\ \ref{res}.

\subsection{Dimensionless numbers}
\label{dn}

% Clean
We consider the capillary radius $r$ and the liquid impact velocity $U_0$ as the characteristic length and velocity, respectively. We choose $U_0$ as the characteristic velocity because (i) it affects flow focusing in the meniscus and (ii) it is the independent variable controlled in the experiments through the voltage applied to the electrodynamic shaker. For fixed values of $L_{\mathrm{bot}}/r$, $L_{\mathrm{mid}}/r$, and $L_{\mathrm{tot}}/r$, the problem can be described in terms of the equilibrium contact angle $\theta_0$, the advancing and receding contact angles, $\theta_a$ and $\theta_r$, and the Bond, Reynolds, and Weber numbers 
\begin{equation}
\text{Bo}=\frac{\rho g r^2}{\sigma}, \quad \text{Re}=\frac{\rho\, U_0\, r}{\mu}, \quad \text{We}=\frac{\rho\, U_0^2\, r}{\sigma},
\end{equation}
where $g$ is gravity. In the absence of surfactant, any dimensionless dependent variable $\Phi$ is a function of the set of dimensionless controlled parameters $\{\theta_0,\theta_a,\theta_r;\text{Bo},\text{Re},\text{We}\}$; i.e., $\Phi=\Phi(\theta_0,\theta_a,\theta_r;\text{Bo},\text{Re},\text{We})$.

% Surfactant
We formally account for the influence of a soluble surfactant by introducing the set of parameters $\{{\cal P}_i\}$, whose values are assumed to depend only on the surfactant molecule; for instance, the Marangoni number, the liquid-air interface sorption constants, the surface and volumetric Peclet numbers, the parameters determining the triple-contact-line dynamics, \ldots. Hereafter, the symbols $\{{\cal P}_i\}_{\textin{Tri}}$ and $\{{\cal P}_i\}_{\textin{Sur}}$ denote $\{{\cal P}_i\}$ evaluated for Triton X-100 and Surfynol 465, respectively. The amount of surfactant dissolved in the droplet is quantified through the ratio $c=\hat{c}/\hat{c}_{\textin{cmc}}$, where $\hat{c}$ is the surfactant concentration and $\hat{c}_{\textin{cmc}}$ is the critical micelle concentration. Therefore, 
\begin{equation}
\Phi=\Phi(\theta_0,\theta_a,\theta_r;\text{Bo},\text{Re},\text{We};\{{\cal P}_i\};c).
\end{equation}

% About the equilibrium surface tension
In the presence of surfactant, we define the Weber number in terms of the equilibrium surface tension $\sigma_0$ rather than the clean-water value, as is common in other experiments. This way, comparing results for different surfactants at the same We reveals the effect of surfactant kinetics rather than the obvious influence of surface tension before jet formation.

% Jet variables
Under our experimental conditions, the problem can be considerably simplified. The smooth inner surface of the capillary is expected to have little contact-angle hysteresis; i.e., $\theta_a\simeq \theta_r$. In addition, $\theta_0$ was practically constant for the liquids considered in our experiments. We conclude that $\theta_0\simeq \theta_a\simeq \theta_r$ were fixed in our experiments and, therefore, they can be neglected in our analysis. In our experiments, the Bond number ranges from 0.03 to 0.05, so we can neglect its effect. Besides, $c$ was constant in our experiments. Therefore, as a first approximation, 
\begin{equation}
\Phi=\Phi(\text{Re},\text{We};\{{\cal P}_i\})
\end{equation}
in the results presented in Sec.\ \ref{res}, where $\Phi$ stands for $V_j/U_0$, $L_j/r$, and $t_jU_0/r$. We will examine the dependence of these quantities on Re, We, and $\{{\cal P}_i\}$ in that section.

\section{Materials and methods}
\label{sec3}

% Microfluidic device
As mentioned in Sec.\ \ref{for1}, Fig.\ \ref{setup1} sketches the device elements used to generate the jet. The main container (A) is a cylindrical borosilicate glass tube with an inner radius of $R=6.9$ mm. A thin borosilicate capillary (B) is positioned concentrically inside the tube. The capillary has an inner radius of $r=0.5$ mm and is located $L_{\mathrm{bot}}=5.12$ mm from the tube bottom. A plug (C) mechanically fixes the capillary to the tube, ensuring both components remain rigidly connected during the impulsive excitation. The plug incorporates two independent ports. One port (D) fills the tube with the working liquid and is sealed afterward. The other port (E) is connected to a syringe pump  ({\sc Legato}210, KD-{\sc Scientific}) to control the pressure inside the capillary. This allows us to fix the heights $L_{\mathrm{tot}}$ and $L_{\mathrm{mid}}$ of the outer and inner meniscus, respectively.  All the experiments were conducted for  $L_{\mathrm{tot}}=2.25$ mm. As mentioned above, we verified that the flow-focusing effect was maximized at $L_{\mathrm{mid}}=0$; that is, for a given $U_0$, the jet velocity was maximum at $L_{\mathrm{mid}}=0$. For this reason, we conducted all experiments under this condition.

\begin{figure*}
\begin{center}
\resizebox{0.65\linewidth}{!}{\includegraphics{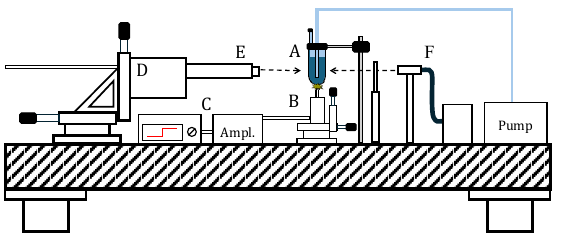}}
\end{center}
\caption{Experimental setup: device (A), electrodynamic shaker (B), step-wave generator and signal amplifier (C), high-speed camera (D), optical lenses (E), and diffuser (F).}
\label{setup2}
\end{figure*}

% Experimental setup
Figure \ref{setup2} shows the experimental setup. We positioned the lower part of the device (A) just above a plastic rod coupled to the electrodynamic shaker (LDS, V101 MACE) (B). A step-wave generator and signal amplifier (LDS, PA25E CE) (C) drove the shaker. We acquired images of the jet with a high-speed camera (FASTCAM SA5) (D) equipped with a telecentric lens (Opto Engineering) (E) at 50000 fps with a magnification of 15 $\mu$m/pixel. We used a backlight illumination system equipped with a diffuser (F). The images were analyzed at pixel resolution.

% Experimental procedure
The shaker applied a controlled impulsive excitation to the bottom of the container. For this purpose, we operated the shaker at 1 Hz and varied the pulse amplitude from 1.05 to 7.5 V. The rod connected to the shaker moved, struck the tube bottom, and generated an upward liquid motion at the instant of impact. We measured the liquid velocity at the tube bottom, $U_0$, as described in the Supplemental Material.   

% Liquids and surfactants
We conducted experiments with ultra-pure Milli-Q water ($\rho=997$ kg/m$^3$ and $\mu=1.0$ mPa$\cdot$s) and aqueous solutions of glycerol (99\%, PanReac AppliChem) at concentrations 19\% (w/w) (W+Gl-19\%), 30\% (w/w) (W+Gl-30\%), and 60\% (w/w) (W+Gl-60\%). Table \ref{tab1} shows the density $\rho$, viscosity $\mu$, and surface tension $\sigma$ of these mixtures. We also tested Surfynol 465 (Evonik) and Triton X-100 (Sigma-Aldrich) dissolved in ultra-pure Milli-Q water. The Surfynol 465 sorption constants have not yet been measured. However, \citet{VSQCC24} showed that the adsorption rate of Surfynol 465 is higher than that of other surfactants at the same relative concentration $\hat{c}/\hat{c}_{\text{cmc}}$, In this sense, Surfynol 465 is a fast-kinetics surfactant. The adsorption characteristic time of Triton X-100 is much longer than that of Surfynol 465 \citep{VSQCC24}. 

\begin{table}
\begin{tabular}{cccc}
& $\rho$ (kg/m$^3$) & $\mu$ (kg/m$\cdot$s) & $\sigma$ (mN/m)\\
\hline 
W & 997 & 0.0010 & 72 \\
\hline 
W+Gl-19\% & 1052 & 0.0015 & 64 \\
\hline 
W+Gl-30\% & 1070 & 0.0020 & 59 \\
\hline
W+Gl-60\% & 1149 & 0.0086 & 56 \\
\hline 
\end{tabular}
\caption{Density $\rho$, viscosity $\mu$, and surface tension $\sigma$ of the liquids used in the experiments.}
\label{tab1}
\end{table}

\begin{figure*}
\begin{center}
\resizebox{0.42\linewidth}{!}{\includegraphics{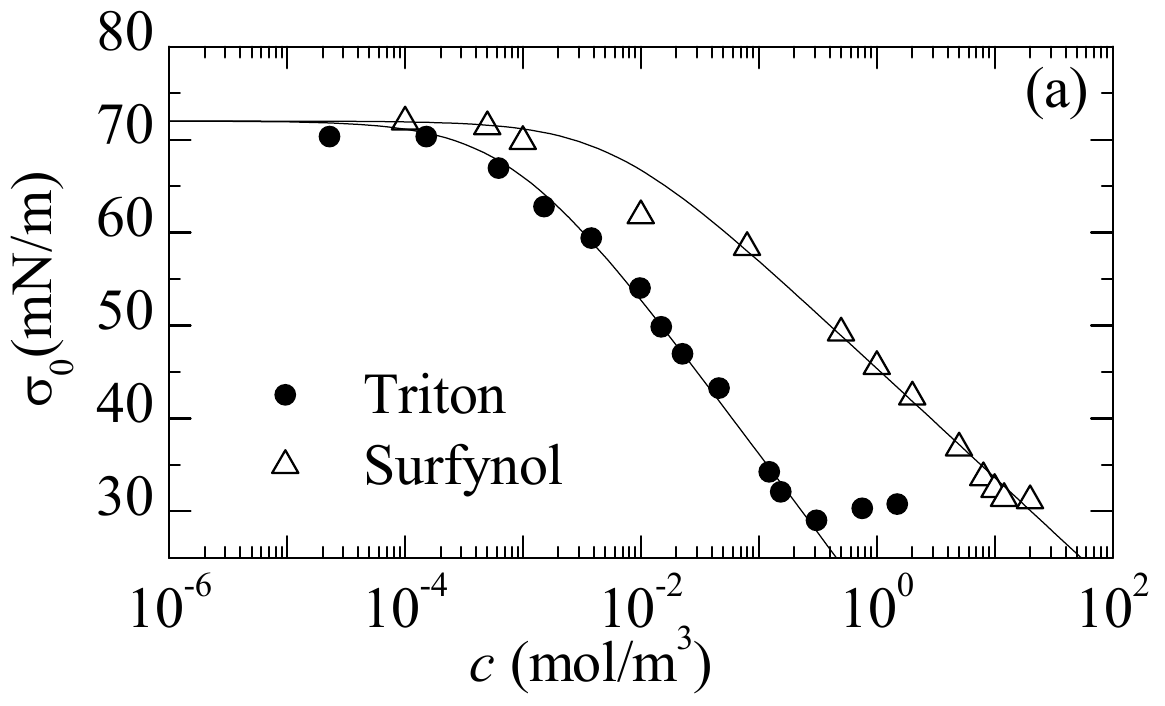}}\resizebox{0.37\linewidth}{!}{\includegraphics{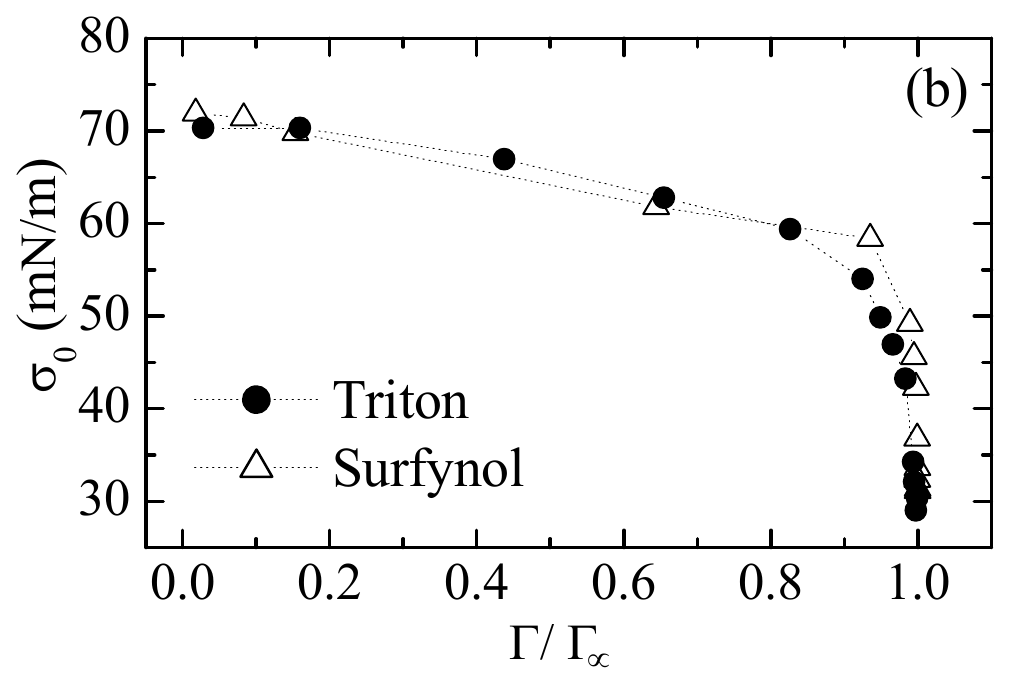}}
\end{center}
\caption{(a) Surface tension $\sigma$ as a function of the surfactant volumetric concentration $\hat{c}$. The lines are the fits of the Langmuir equation of state to the Surfynol 465 and Triton X-100 data. (b) Surface tension $\sigma$ as a function of the normalized surfactant surface concentration $\Gamma/\Gamma_{\infty}$. The arrows indicate the concentrations considered in the experiments.}
\label{st}
\end{figure*}

% Surface tensions. Surfynol 465 and Triton
Figure \ref{st}a shows the surface tension $\sigma$ as a function of the surfactant volumetric concentration $\hat{c}$ for Surfynol 465 and Triton X-100 measured in our laboratory using a drop-shape tensiometer \citep{M24}. The critical micelle concentrations for Surfynol 465 and Triton X-100 are $\hat{c}_{\text{cmc}}=10$ and 0.24 mM, respectively. The Langmuir equation of state
\begin{equation}
\sigma=\sigma_c-\Gamma_{\infty} R_g T\log(1+L_d\, \hat{c}/\Gamma_{\infty})
\end{equation}
was fitted to those values. Here, $\sigma_c=72$ mN/m is the water-air surface tension, $\Gamma_{\infty}$ is the maximum packing density, $R_g$ is the gas constant, $T$ is the temperature, and $L_d$ is the depletion length. In the case of Surfynol 465, the value $\Gamma_{\infty}=2.1$ $\mu$mol/m$^2$ was taken from \citet{PZPA16}, and $L_d=0.38$ mm was obtained from the fit. For Triton X-100, the fit yielded $\Gamma_{\infty}=3.03$ $\mu$mol/m$^2$ and $L_d=3.83$ mm. Using these values, the Langmuir adsorption isotherm
\begin{equation}
\label{km2}
\frac{\Gamma/\Gamma_{\infty}}{1-\Gamma/\Gamma_{\infty}}=\frac{L_d\,  \hat{c}}{\Gamma_{\infty}}
\end{equation}
allows one to obtain $\sigma(\Gamma)$ from $\sigma(\hat{c})$ (Fig.\ \ref{st}b), where $\Gamma$ is the surfactant surface concentration. As shown, $\sigma(\Gamma)$ is similar for the two surfactants. The differences between their effects on the jet generation can be mainly attributed to their different sorption kinetics. 

% Concentrations
We conducted experiments with Surfynol 465 and Triton X-100 at the same relative concentration $c=\hat{c}/\hat{c}_{\textin{cmc}}=2$, so that the equilibrium surface tension was practically the same for the two surfactants ($\sigma=33$ and 31 mN/m for Surfynol 465 and Triton X-100, respectively). 

\begin{figure}
\begin{center}
\resizebox{0.47\textwidth}{!}{\includegraphics{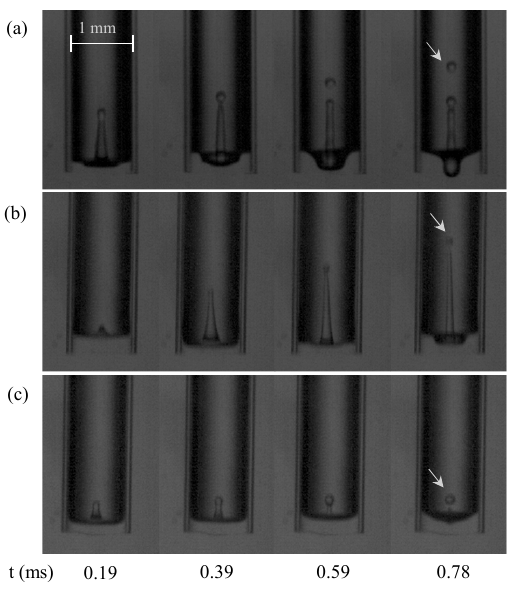}}\
\end{center}
\caption{Sequence of images obtained for W+Gl-19\% (a) and water loaded with Surfynol 465 (b) and Triton X-100 (c) at $\text{Re}=18.7\pm 0.09$ and $\text{We}=0.020\pm 0.001$. The Surfynol 465 and Triton X-100 concentrations were $c/c_{\textin{cmc}}=2$. The arrow points to the first-emitted droplet. Adding Surfynol 465 further accelerates jet emission, eventually exceeding the speed attained by W+Gl-19\%, a clean interface with the same reduced equilibrium surface tension. Conversely, Triton X-100 partially suppresses the jet formation.}
\label{images}
\end{figure}

For illustration, Fig.\ \ref{images} shows the sequence of images obtained for W+Gl-19\% and water loaded with Surfynol 465 and Triton X-100. Without surfactant, a cavity forms at the jet base, disrupting the flow-focusing mechanism. This cavity disappears with the surfactant, allowing the Surfynol 465 jet to overtake the W+Gl-19\% jet. We will return to this in Sec.\ \ref{res}.

\section{Results}
\label{res}

\subsection{Results without surfactants}

% Introduction
We first show the results obtained without surfactants. We measured the jet formation time $t_j$ (the time between jet inception and droplet detachment), the jet length $L_j$ just before droplet detachment, and the average velocity $V_j=L_j/t_j$. The equilibrium contact angle $\theta_0$ ranged from $45^{\circ}$ to $49^{\circ}$, allowing us to focus on the effects of We and Re, as explained in Sec.\ \ref{dn}. 

\begin{figure}[tbp]
\begin{center}
\resizebox{0.9\linewidth}{!}{\includegraphics{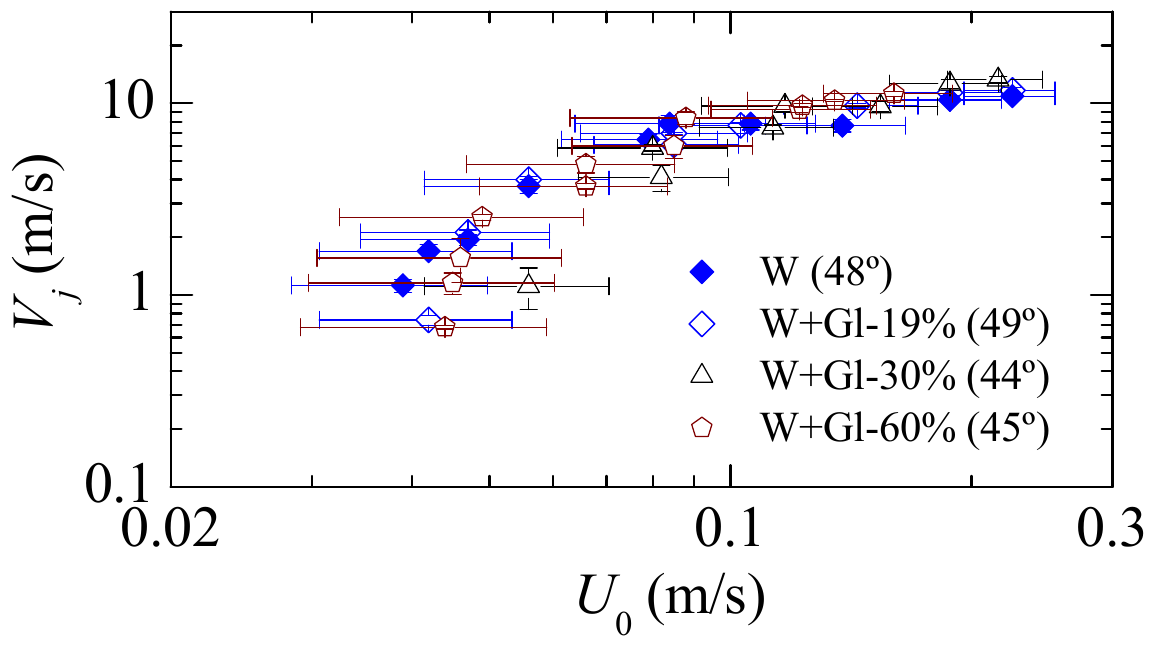}}
\end{center}
\caption{Jet velocity $V_j$, length $L_j$, and time $t_j$ as a function of liquid velocity $U_0$ after the impact. The numbers in parentheses indicate the meniscus contact angle.}
\label{U0volt}
\end{figure}

% Dimensional quantities
First, we examine the jet velocity $V_j$ as a function of liquid velocity $U_0$ after impact (Fig.\ \ref{U0volt}). The flow-focusing effect increases jet velocities by up to two orders of magnitude over $U_0$. For sufficiently large $U_0$, the jet velocity is essentially independent of liquid properties, underscoring the secondary role of viscosity and surface tension in this inertial regime. Significant differences arise for small $U_0$. Viscosity and surface tension reduce $V_j$. As glycerine concentration increases, viscosity increases and surface tension decreases, explaining the non-monotonic behavior of $V_j$ with glycerine concentration.

\begin{figure*}[tbp]
\begin{center}
\resizebox{0.408\textwidth}{!}{\includegraphics{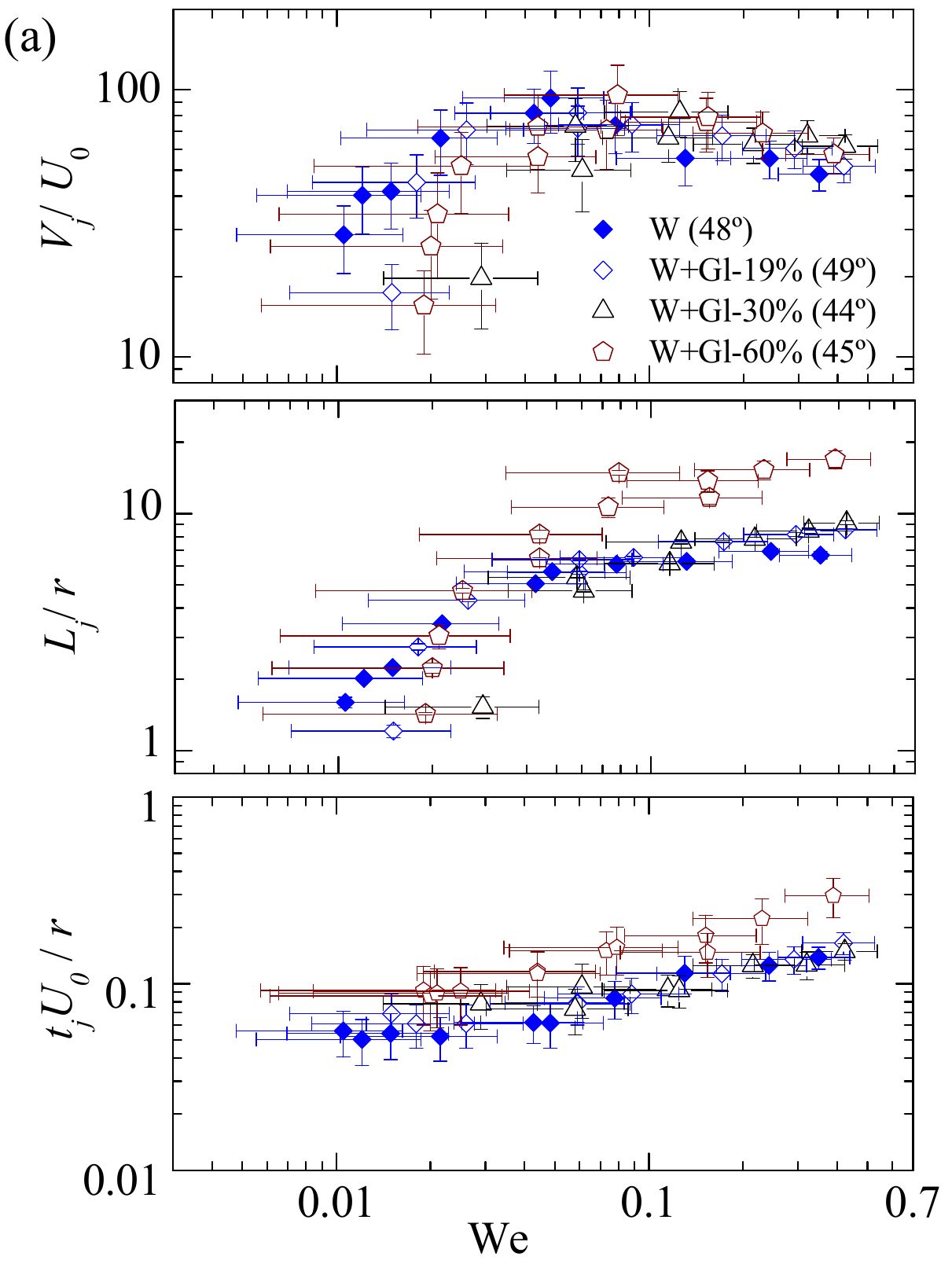}}\resizebox{0.39\textwidth}{!}{\includegraphics{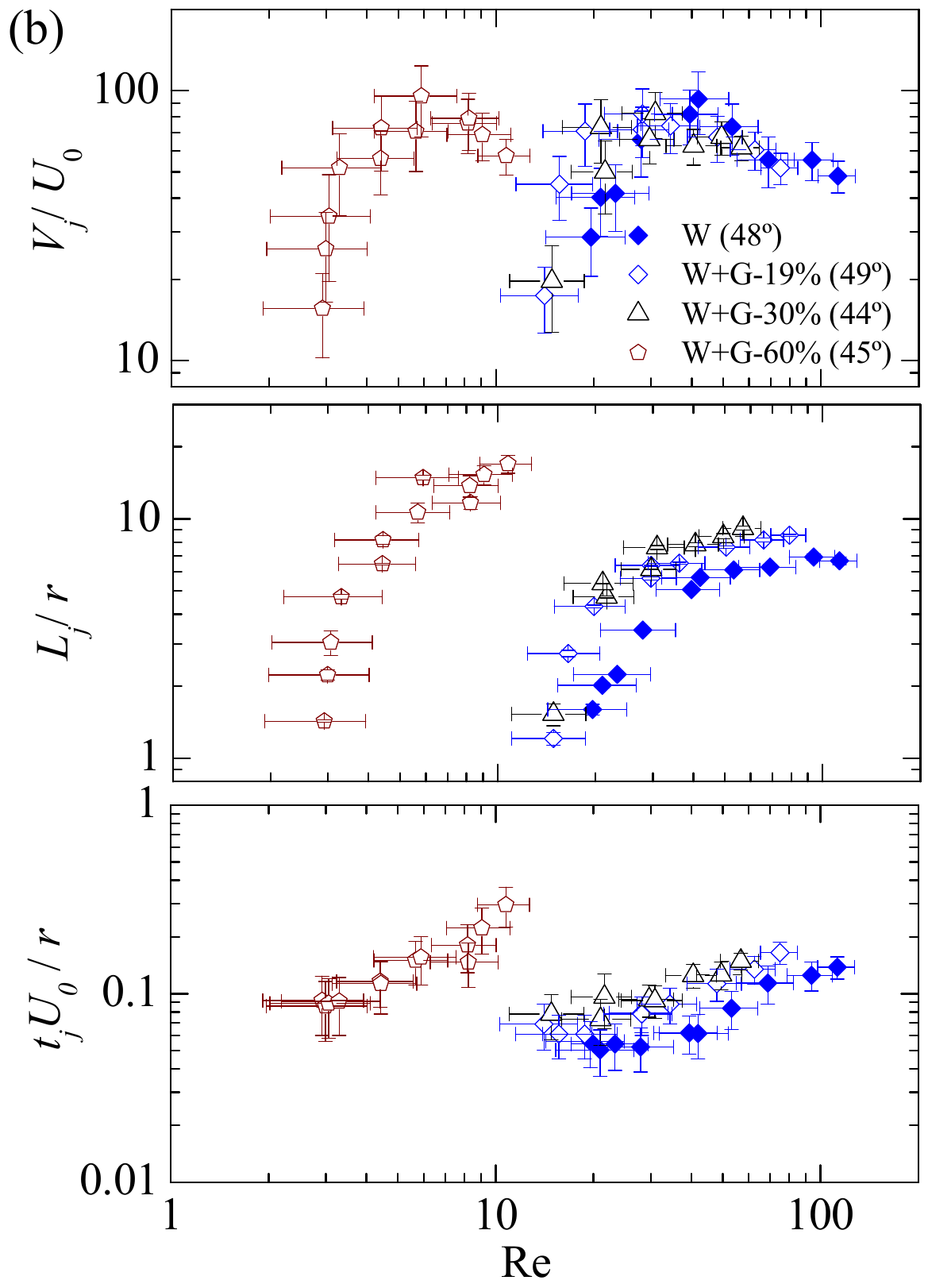}} 
\end{center}
\caption{Dimensionless jet velocity $v_j/U_0$, length $l_j/r$, and emission time $t_j/(r/U_0)$ as a function of the Weber (a) and Reynolds (b) numbers. The numbers in parentheses indicate the meniscus contact angle.}% The Ohnesorge number (Oh) is 0.005 for pure water, 0.008 for the 19\% glycerol–water mixture, and 0.011 for the 30\% glycerol–water mixture.}
\label{WeRe}
\end{figure*}

% Dimensionless quantities
Figure \ref{WeRe} shows the results as a function of We and Re. The data collapse much better when plotted against the Weber number, indicating that viscosity plays a secondary role for the range considered in our experiments. Specifically, the same values of $V_j/U_0$, $L_j/r$, and $t_j U_0/r$ are obtained even when the Reynolds number changes by one order of magnitude. 
 
% Viscosity effect
The scattering of the curves $\Phi(\text{We})$ in Fig.\ \ref{WeRe}a is essentially due to the secondary effect of viscosity. The viscous boundary layer grows from the impact. The boundary layer thickness $\delta$ at jet formation is $\delta\sim \sqrt{\nu\, \Delta t}$, where $\nu$ is the kinematic viscosity and $\Delta t$ is the time elapsed until the jet forms \citep{schlichting1979}. In our experiments, $\delta/r\sim 0.1$. Therefore, viscous effects on flow focusing are not negligible. In addition, the high viscosity of W+G-60\% delays jet breakup ($t_j$ increases), allowing the jet to reach larger $L_j$ without increasing velocity.  

% The maximum
The ratio $V_j/U_0$ measures the momentum focused on (transferred to) the jet. Interestingly, $V_j/U_0$ depends non-monotonically on both the Weber and Reynolds numbers (Fig.\ \ref{WeRe}). This indicates an optimum $U_0$ that maximizes the momentum transferred to the jet. This result can be rationalized as follows. Surface tension is a momentum sink associated with the interface deformation, which explains why jet emission ceases below a threshold Weber number. This momentum sink is essentially independent of $U_0$. Therefore, its effect on $V_j/U_0$ decreases as $U_0$ (We) increases. This explains the increase in the momentum transferred to the jet, $V_j/U_0$, as We increases. For sufficiently large We, surface tension becomes irrelevant, and $V_j/U_0$ is expected to tend toward a constant value (the inertial regime). However, a significant portion of the momentum injected into the liquid is consumed by the meniscus's upward motion during jet formation, affecting the flow focusing, as explained in Sec.\ \ref{im}. This effect increases with $U_0$, reducing the ratio $V_j/U_0$ for large We ($U_0$). Together, surface tension and meniscus motion produce a maximum in the curve $V_j/U_0(\text{We})$.

\begin{figure}[tbp]
\begin{center}
\resizebox{0.45\textwidth}{!}{\includegraphics{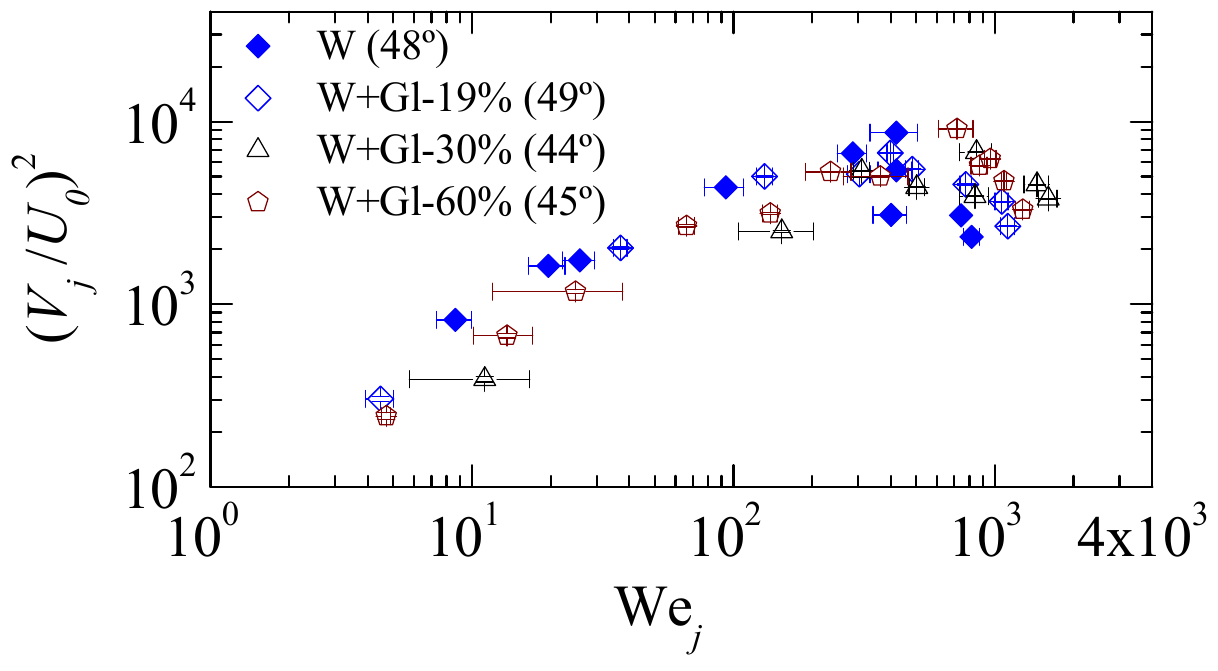}}
\end{center}
\caption{Kinetic energy density transferred to the jet, $(V_j/U_0)^2$, as a function of the Weber number based on the jet velocity, $\text{We}_j=\rho V_j^2 r/\sigma$. The numbers in parentheses indicate the meniscus contact angle.}
\label{new}
\end{figure}

% Comment Weber number
The Weber number is low in all experiments because we chose the bottom liquid velocity $U_0$ as the characteristic velocity. We chose this quantity because it affects the flow focusing in the meniscus, as explained above, and is the independent variable controlled in the experiments through the voltage applied to the electrodynamic shaker. However, the liquid velocity increases sharply in the jet, reaching values up to two orders of magnitude larger than $U_0$. If we define the Weber number in terms of $V_j$ instead of $U_0$, $\text{We}_j=\rho V_j^2 r/\sigma$, the transition to the inertial regime occurs for $\text{We}\gtrsim 10^3$ (Fig.\ \ref{new}). As occurs with We, an optimum Weber number We$_j$ maximizes the kinetic energy density transferred to the jet, $(V_j/U_0)^2$.

\subsection{Results with surfactants}

\begin{figure}
\begin{center}
\resizebox{0.37\textwidth}{!}{\includegraphics{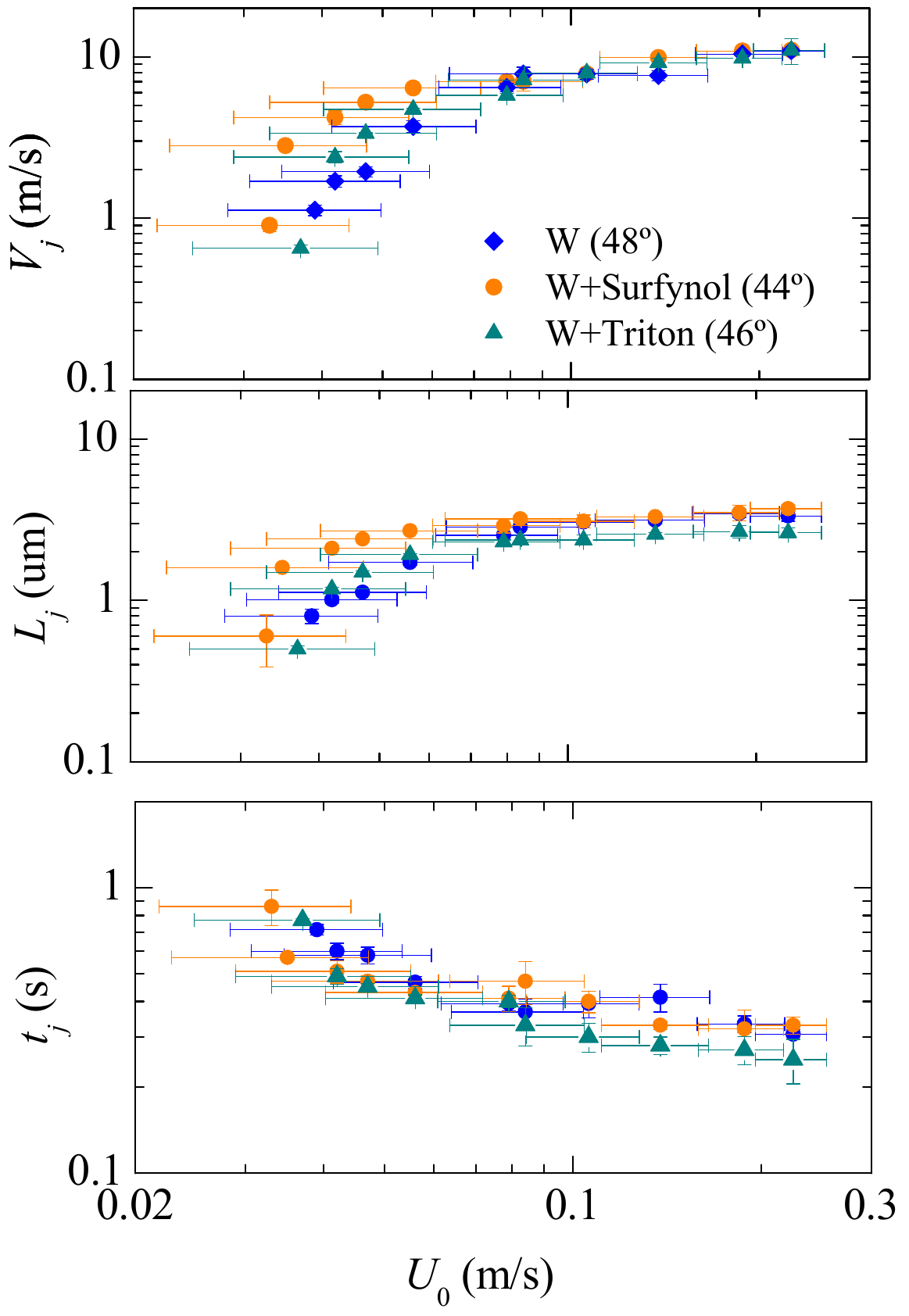}} 
\end{center}
\caption{Jet velocity $V_j$, length $L_j$ and emission time $t_j$ as a function of liquid velocity $U_0$ after the impact. The numbers in parentheses indicate the meniscus contact angle.}
\label{sur_u0}
\end{figure}

% Adding surfactants to water
Figure \ref{sur_u0} shows the effect of adding the surfactant to water. Surfynol 465 reduces the $U_0$ threshold for emission. At small and moderate values of $U_0$, the solution containing Surfynol 465 exhibits higher jet velocities and longer jets than water. This behavior can be attributed to Surfynol 465's fast adsorption kinetics, which allow it to restore equilibrium surface tension during interface expansion and thus favor jet emission. Triton X-100's effect is negligible because of its slower adsorption kinetics. Because surfactant is present at 2 CMC, a significant fraction exists as micelles in the bulk liquid. Triton X-100 is transported toward the newly created interface as both monomers and micelles that disintegrate near the interface. Under rapid interface expansion, Triton X-100 adsorption may be insufficient to maintain homogeneous interfacial coverage. This could result in non-uniform surfactant distribution and surface-tension gradient, particularly around the jet base. The resulting Marangoni stresses may oppose interfacial motion, effectively increasing interfacial resistance to deformation (interfacial immobilization) and counteracting the surface-tension-reduction effect. 

At high $U_0$, the jet characteristics of all fluids progressively converge because inertia dominates the flow and the surfactant has a negligible effect. In this inertial regime, the jet velocity increases slightly with $U_0$, while the emission time slightly decreases, resulting in a practically constant jet length.

\begin{figure*}[tbp]
\begin{center}
\resizebox{0.4\textwidth}{!}{\includegraphics{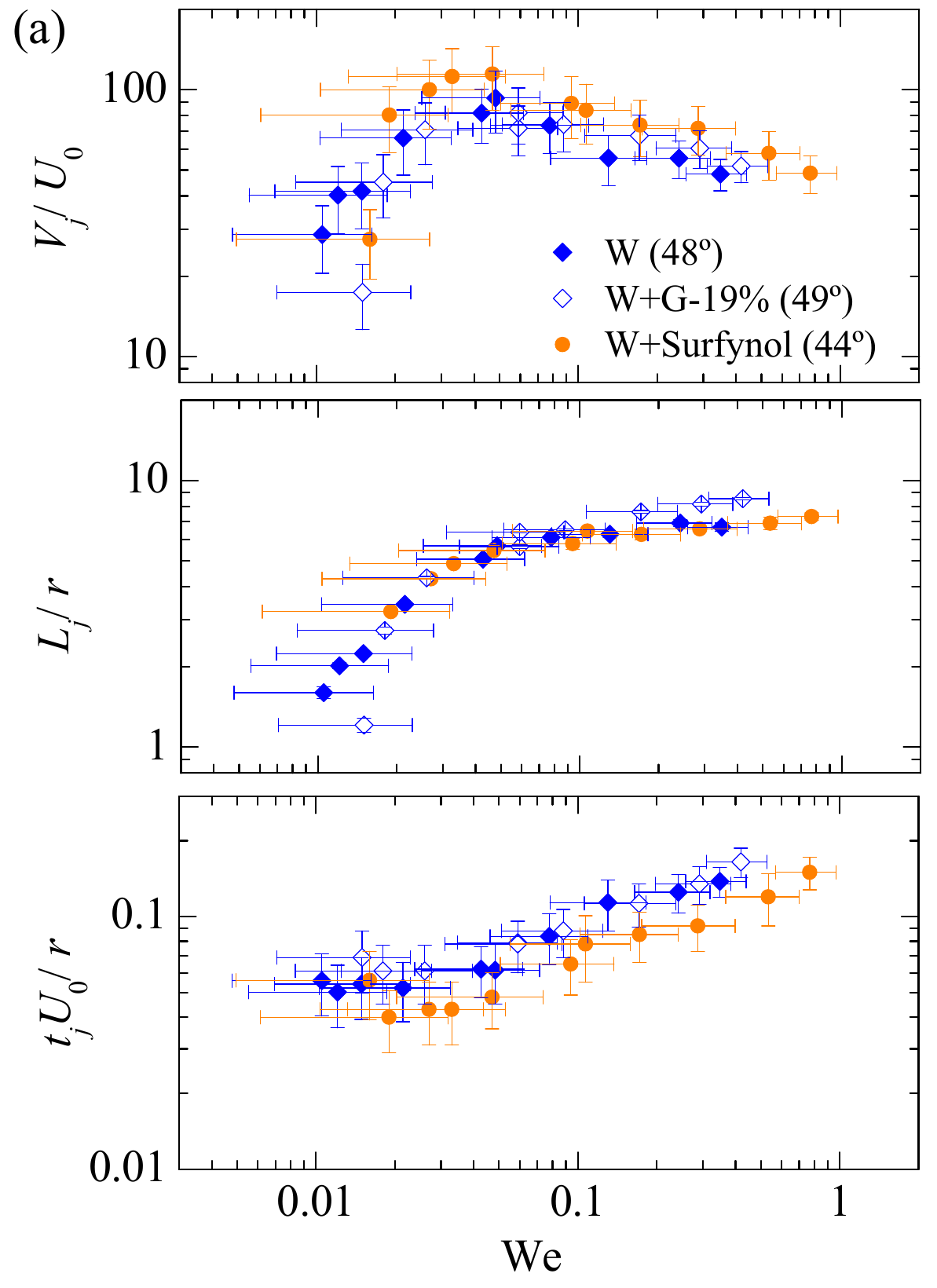}} \resizebox{0.4\textwidth}{!}{\includegraphics{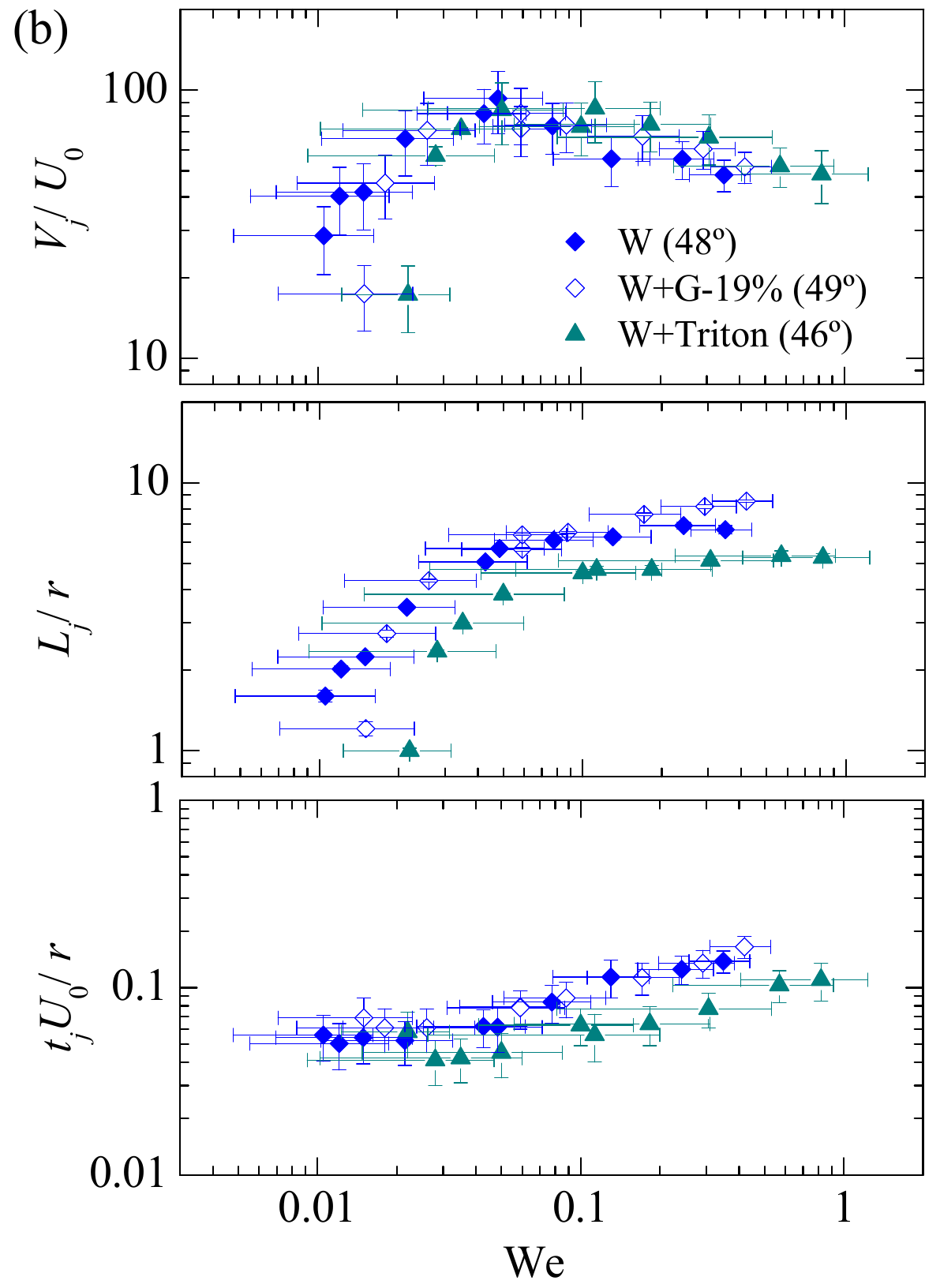}}
\end{center}
\caption{Dimensionless jet velocity \(V_j/U_0\), length \(l_j/r\), and emission time \(t_j/(r/U_0)\) as a function of the Weber number We, for \(Oh = 0.005\) (water) and \(0.008\) (W+Gl-19\% and water with Surfynol 465 (a) and Triton X-100 (b)). The numbers in parentheses indicate the meniscus contact angle.}
\label{sur0}
\end{figure*}

% Ohnesorge number. Water
Dimensional analysis helps us understand the physical mechanisms associated with the surfactants. For this purpose, it is convenient to consider the Ohnesorge number $\text{Oh}=\text{We}^{1/2}/\text{Re}=\mu/(\rho r \sigma_0)^{1/2}$ instead of Re ($\sigma_0$ is the equilibrium surface tension). Figure \ref{sur0}a compares the dimensionless jet velocity, length, and emission time of water and water loaded with Surfynol 465. Surfynol 465 increases the Ohnesorge number (decreases the equilibrium surface tension). Consequently, it increases the Weber number threshold for emission and decreases $V_j/U_0$ for small Weber numbers.  

% W+Gl-19\% and Surfynol
We now consider the mixture W+Gl-19\% whose physical properties yield an Ohnesorge number $\text{Oh}\simeq 0.008$ similar to that of water loaded with Surfynol 465 or Triton X-100. Interestingly, $V_j/U_0$ for Surfynol 465 was larger than for W+Gl-19\% for the same Weber number. This suggests that Surfynol 465 enhances jet emission beyond the acceleration achieved for a clean interface with the same reduced equilibrium surface tension. This can be explained by the fast adsorption of Surfynol 465, which enables quasi-instantaneous replenishment of the interface created during jet growth \citep{FVMGF26}. Thus, surface tension remains nearly at equilibrium during most of the process, making the rigidifying Marangoni stress negligible. In addition, Surfynol 465 enhances the energy focusing at the jet base, leading to even higher values of $V_j/U_0$.

% The cavity
Without Surfynol 465, a cavity forms at the jet base, disrupting the flow-focusing mechanism  (Fig.\ \ref{images}). With Surfynol 465, this cavity disappears or appears later, allowing the Surfynol 465 jet to overtake the W+Gl-19\% jet (Fig.\ \ref{images}). Surfynol 465 adsorption may not fully replenish the interface at the beginning of the jet formation because the characteristic time is very short, on the order of, or shorter than, $10^{-4}$ s, as discussed below. Marangoni stress may arise during jet birth from an uneven distribution of surfactant over the interface, which may immobilize it and delay jet growth. However, Marangoni stress may prevent the cavity formation enhancing the flow focusing effect. As time passes, the equilibrium surface tension is restored, the jet accelerates, and a cavity forms at this base, as occurs for W+Gl-19\%  (Fig.\ \ref{images}). An alternative explanation for the surfactant cavity suppression relies on surfactant influence on triple-contact-line motion.

% W+Gl-19\% and Triton
In the case of Triton X-100 (Fig.\ \ref{sur0}b), the Weber number threshold for emission is larger than that of W+Gl-19\%. In addition, $V_j/U_0$ and $L_j/r$ are smaller than those for W+Gl-19\% at small Weber numbers (see Fig.\ \ref{images}). This reflects Triton X-100's inability to restore equilibrium surface tension, which generates Marangoni stress that partially immobilizes the interface. Marangoni stress becomes less important in the inertial regime. This, along with enhanced flow focusing, allows the jet to reach velocities slightly larger than those of W+Gl-19\% at large We. The difference between $\Phi=\Phi(\text{We})$ for Surfynol 465 and Triton X-100 corresponds to the effect of $\{{\cal P}_i\}$. Since both surfactants have similar Marangoni numbers (see Sec.\ \ref{sec3}), this effect likely arises from surfactant kinetics.

\begin{figure*}[tbp]
\begin{center}
\resizebox{0.4\textwidth}{!}{\includegraphics{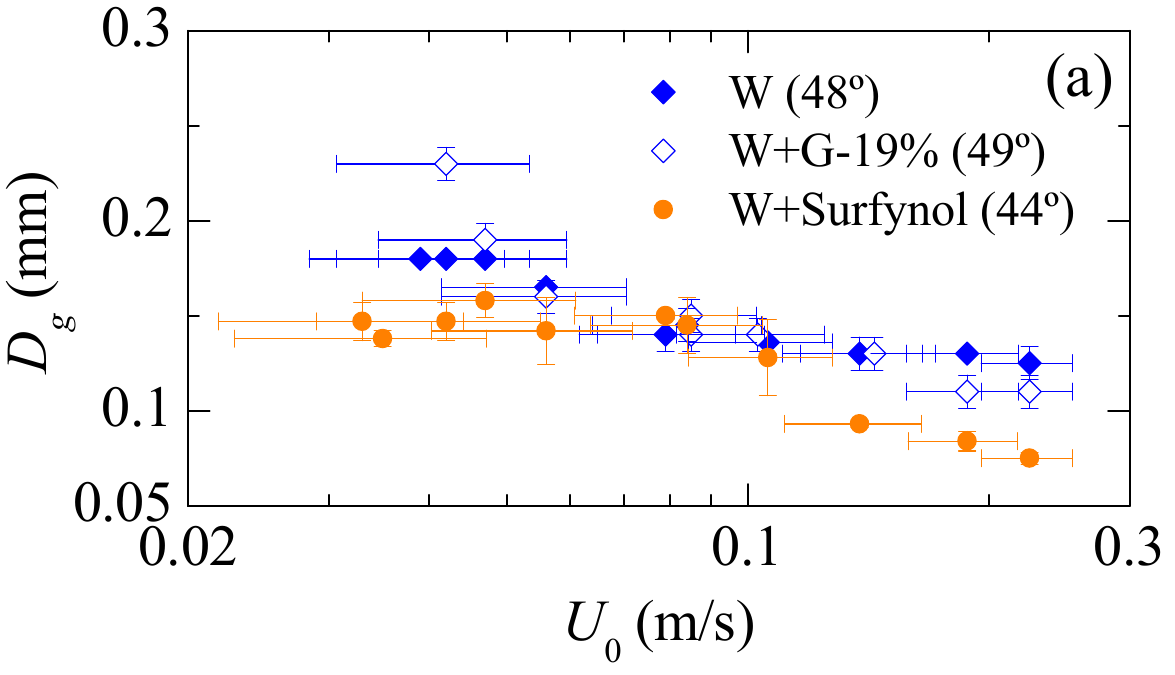}} \resizebox{0.4\textwidth}{!}{\includegraphics{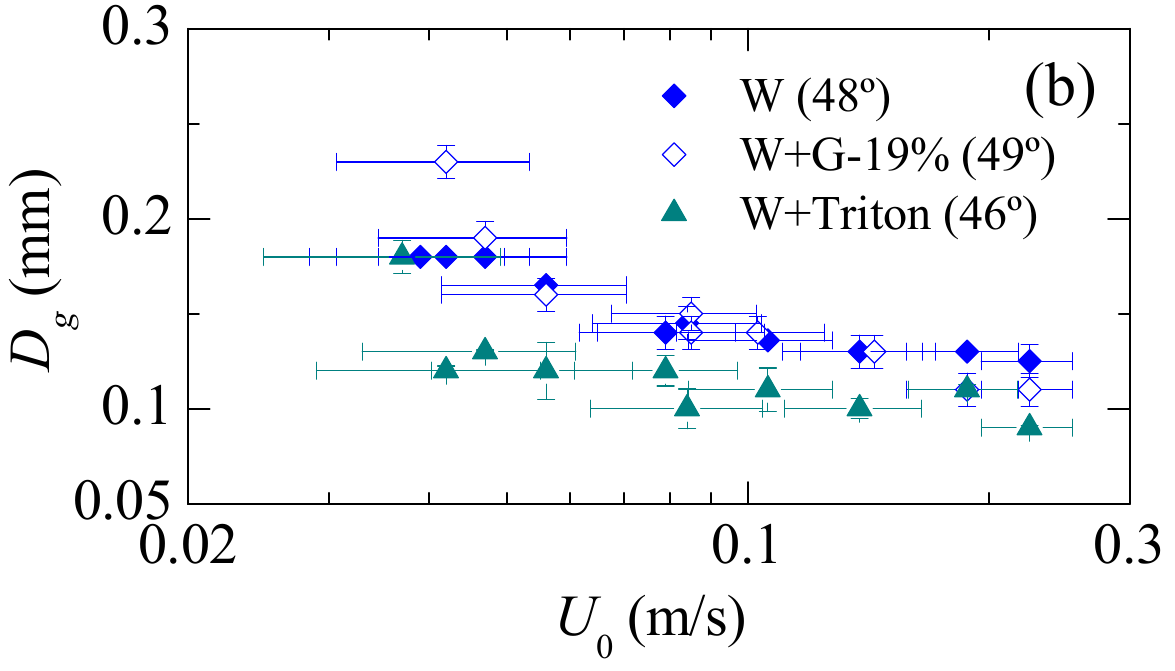}}
\end{center}
\caption{Droplet diameter $D_g$ as a function of liquid velocity $U_0$ after the impact. The numbers in parentheses indicate the meniscus contact angle. }
\label{surX}
\end{figure*}

% Droplet diameter
Figure \ref{surX} shows the surfactant's influence on the first-emitted droplet diameter $D_g$. Both surfactants reduce the jet diameter and, consequently, the droplet size across the impact-velocity range analyzed in our experiments. At the highest velocity $U_0$, Surfynol 465 reduces the droplet diameter by 40\%, which may benefit technological applications such as printing technology.  This effect can be explained by convection of surfactant molecules toward the jet tip in the early stage of its formation, which does not rely on surfactant adsorption. This is analogous to what occurs in the viscous entrainment mode of the selective withdrawal configuration, where surfactant produces much thinner liquid threads for the same dimensionless parameters \citep{RM23}.

\begin{figure}[tbp]
\begin{center}
\resizebox{0.4\textwidth}{!}{\includegraphics{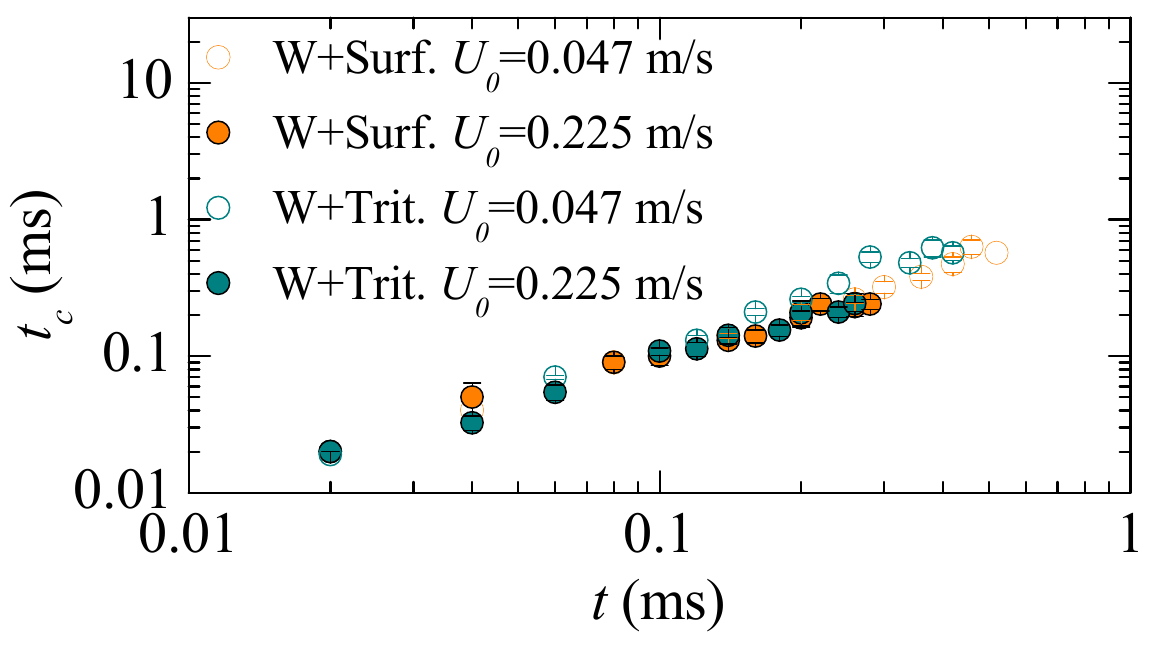}} 
\end{center}
\caption{Characteristic time $t_c$ as a function of time $t$ for several impact velocities $U_0$.}
\label{tc_t}
\end{figure}

% Characteristic times
Consider a growing jet of an approximately constant radius $R_j$ and an increasing length $\hat{L}_j(t)$. The relative variation of the free surface area, $S_j(t)\approx \pi R_j^2 \hat{L}_j$, is $S_j^{-1} dS/dt\approx \hat{L}_j^{-1}d\hat{L}_j/dt=d\ln \hat{L}_j/dt$. The inverse of this quantity, $t_c=(d\ln \hat{L}_j/dt)^{-1}$ can be identified as the {\em instantaneous} hydrodynamic time. Figure \ref{tc_t} shows that $t_c$ ranges from $\sim 10$ $\mu$s to $\sim 1$ ms. The interface grows faster at the beginning of jet formation, when surfactant kinetics are expected to be more relevant. As anticipated above, Marangoni stress is expected to arise in the jet base for both Surfynol 465 and Triton X-100.

\section{Conclusions}

We experimentally studied impulsive jet generation by flow focusing. We considered the singular case in which the liquid meniscus is at the bottom of the capillary, where the flow-focusing effect is maximized. This effect increases jet velocities by up to two orders of magnitude over the impact liquid velocity $U_0$. For sufficiently large $U_0$, the surfactant-free jet velocity is essentially independent of liquid properties, underscoring the secondary role of viscosity and surface tension in this inertial regime. For each liquid, the competition of surface tension momentum sink with meniscus dynamics yields an optimum $U_0$ that maximizes momentum transfer to the jet.

Adding surfactant to water lets the liquid spread over the inner capillary surface during jet formation. Surfynol 465 reduces the $U_0$ threshold for emission and increases jet velocity at small $U_0$ because it rapidly restores equilibrium surface tension during interface growth, favoring emission. Triton X-100 has a smaller effect because of its slower kinetics. For sufficiently large values of $U_0$, the flow is dominated by inertia and both surfactants have a negligible effect. Both surfactants reduce the jet diameter and, consequently, the droplet size across the impact-velocity range analyzed in our experiments. At the highest impact velocity, $U_0$, Surfynol 465 reduces the droplet diameter by about 40\%, which may be advantageous for technological applications such as printing.

We compared the aqueous surfactant solutions with a glycerine-water mixture that yielded a nearly identical Ohnesorge number. The jet loaded with Surfynol 465 was even faster than the glycerine-water jet at the same Weber number. This implies that adding Surfynol 465 accelerates jet emission even beyond the speed of a clean interface with the reduced equilibrium surface tension.

Our results show that adding a surfactant can delay or even suppress the cavity at the bottom jet (Fig.\ \ref{images}), enhancing the flow-focusing mechanism and leading to faster jet growth with Surfynol 465. An open question is why the two surfactants produce this effect. Surfactant adsorption may not fully replenish the cavity interface because the characteristic time is very short, on the order of, or shorter than, $10^{-4}$ s. Marangoni stress arising from an uneven distribution of surfactant over the interface may immobilize it, precluding cavity growth at the beginning of jet formation. This may occur even with a fast-kinetic surfactant such as Surfynol 465, for which the cavity formation is delayed (Fig.\ \ref{images}). An alternative explanation relies on surfactant influence on triple-contact-line motion, which may affect the cavity formation. 

\textbf{Acknowledgements}
Support from the Spanish Ministry of Science and Innovation (MCIN) through grant PID2025-171659NB-I00/ AEI// FEDER, UE is gratefully acknowledged. This work was co-funded (85\%) by the European Union through the European Regional Development Fund (FEDER) and the Regional Government of Extremadura. Managing Authority: Ministry of Finance (Grant GR24077).

\bibliography{central,central2,addref,references}

\end{document}